\documentclass[showpacs,twocolumn,floats,superscriptaddress]{revtex4}
\usepackage{graphicx}
\usepackage{amsmath}
\usepackage{bm}

\newcommand{\bq}{\begin{equation}}
\newcommand{\ee}{\end{equation}}
\newcommand{\fr}[2]{\frac{#1}{#2}}
\newcommand{\eps}{\varepsilon}

\newcommand{\ve}{\vec e}

\newcommand{\vj}{\vec j}
\newcommand{\vn}{\vec n}

\newcommand{\vp}{\vec p}
\renewcommand{\vr}{\vec r}

\newcommand{\vs}{\vec s}

\newcommand{\vA}{\vec A}

\newcommand{\vK}{\vec K}

\newcommand{\vsigma}{\mbox{\boldmath $\sigma$}}
\newcommand{\vtau}{\mbox{\boldmath $\tau$}}

\renewcommand{\vec}[1]{\mathbf{#1}}

\begin{document}
\title{Spin and Charge Structure of the Surface States in Topological Insulators
%Pseudo-Helical Surface States in Topological Insulators
%\\ "How Helical are the Surface States in Topological Insulators?"
}

\date{\today }

\author{P. G. Silvestrov}
\author{P. W. Brouwer}
\affiliation{Physics Department and Dahlem Center for Complex
Quantum Systems, Freie Universit\"{a}t Berlin, Arnimallee 14,
14195 Berlin, Germany}
\author{E. G. Mishchenko}
\affiliation{Department of Physics and Astronomy, University of
Utah, Salt Lake City, UT 84112}

\begin{abstract}

We investigate the spin and charge densities of surface states of
the three-dimensional topological insulator $Bi_2Se_3$, starting
from the continuum description of the material [Zhang {\em et
al.}, Nat.\ Phys.\ {\bf 5}, 438 (2009)]. The spin structure on
surfaces other than the $(111)$ surface has additional complexity
because of a misalignment of the contributions coming from the two
sublattices of the crystal. For these surfaces we expect new
features to be seen in the spin-resolved ARPES experiments, caused
by a non-helical spin-polarization of electrons at the individual
sublattices as well as by the interference of the electron waves
emitted coherently from two sublattices. We also show that the
position of the Dirac crossing in spectrum of surface states
depends on the orientation of the interface. This leads to contact
potentials and surface charge redistribution at edges between
different facets of the crystal.

\end{abstract}
% \pacs{73.63.-b,
%81.05.Uw, 73.43.-f}
 \maketitle

%{\it Which pictures to draw?}

\section{Introduction} \label{sec:1}

%{\em 1.Introduction.}
A new class of materials, topological insulators,
%(TIs),
have attracted a great deal of attention in the last two years
(see Refs. \onlinecite{HasanKaneReview10,ShouChengZhangReview11}
for reviews). Typically, these are band insulators for which
strong spin-orbit coupling leads to an inversion of the bulk band
gap. The band inversion~\cite{VolkovPankratov85} causes the
emergence of protected two-dimensional states on the surface of
this in other respects conventional bulk insulator.

The surface states of a topological insulator have a conical
energy spectrum, characteristic of eigenmodes of the massless
Dirac equation. They are routinely described by the effective
two-dimensional Hamiltonian~\cite{Fu09}
 \bq
  \label{surfaceHamiltonian}
  H_S=v_F (\sigma_x p_y -\sigma_y p_x),
 \ee
where $v_{\rm F}$ is the Fermi velocity and $\sigma_x$ and
$\sigma_y$ are Pauli matrices. The operator ${\bm
\sigma}=(\sigma_x, \sigma_y, \sigma_z)$
%in the effective surface Hamiltonian (1)
is usually identified with the true electron spin~\cite{examples}.
The spin polarization is then perpendicular to the electron's
momentum, which leads to the surface states being referred to as
``helical''~\cite{endnote}. Yet, we are not aware of a detailed
investigation of the relation between ${\bm \sigma}$ and the true
spin. The nature of ${\bm \sigma}$ is not important for a number
of phenomena involving the surface states, in particular to those
that only depend on the nontrivial Berry phase acquired by
surface-state electrons~\cite{AharonovBohm,Bardarson2010}.
However, for the properties that explicitly probe the electron's
spin the nature of operators in the effective Hamiltonian becomes
of principal importance \cite{FW2,Lasia2012}. In this paper we
elucidate how the spin structure of the surface electrons follows
from the bulk Hamiltonian of the topological insulator and show
that the relation between spin and momentum is richer and more
delicate than suggested by a naive interpretation of Eq.\
(\ref{surfaceHamiltonian}).

\begin{figure}
\includegraphics[width=8cm]{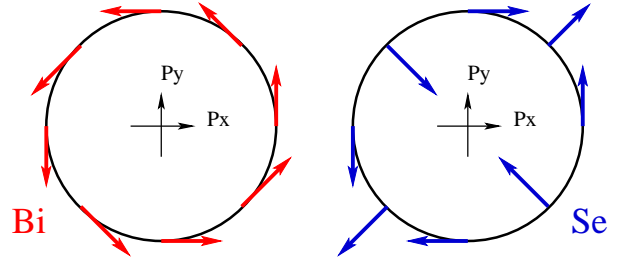}
\caption{\label{fig:1} Polarization of a surface-state electrons
on the \mbox{Bi} and \mbox{Se} sublattices for different
orientation of the in-plane momentum $p$ for a surface orthogonal
to the quintuple layer. Spin textures for a symmetric model
Eq.~(\ref{HamSymmetric}) are shown to demonstrate the inevitable
strong difference between the two sublattices spin for surfaces
other than $(111)$. Details of the spin behavior for the
asymmetric case are given in Sec.~\ref{sec:4}.
%and polarization of the final
%photoexcited electron.
}
%\vspace{-.3cm}
\end{figure}

As a specific material we will consider the compound Bi$_2$Se$_3$,
presently probably the most popular example of a topological
insulator. This material was considered both theoretically and
experimentally in Refs.~\onlinecite{HasanBiSe,ZhangBiSe}, where it
was shown that Bi$_2$Se$_3$ has a large topologically nontrivial
band gap $\sim 0.3$ eV, leading to surface states with a single
Dirac cone. Zhang
%{\em et al.}
and coworkers \cite{ZhangBiSe,ModelHam10} also suggested a simple
three-dimensional Hamiltonian describing the long-wavelength
electronic dynamics in Bi$_2$Se$_3$. This low-energy Hamiltonian
%of Ref.~\onlinecite{ZhangBiSe}
contains both the true electron spin and a pseudospin as its
primary degrees of freedom, where the pseudospin refers to states
with support on the \mbox{Bi} and \mbox{Se} sublattices.

Bi$_2$Se$_3$ is a strongly anisotropic material, with a layered
structure involving quintuple layers of Bi and Se atoms. (One
quintuple layer consists of three Se layers strongly bonded to two
Bi layers in between.) Yet, by virtue of their topological
protection, the surface states exist for arbitrarily oriented
crystal surfaces, not only at the $(111)$ surface parallel to the
quintuple layer. Based on the low-energy continuum Hamiltonian of
Zhang {\em et al.}~\cite{ZhangBiSe}, we here show that the
$\vsigma$-operator entering the surface Hamiltonian
Eq.~(\ref{surfaceHamiltonian}) coincides with the electron's spin
for the $(111)$ surface only. For any other surface orientation,
the spin content of the electron wave function components is
different on different sublattices, see Fig.~1. Whereas most
experiments are carried out for the $(111)$ surface of
Bi$_2$Se$_3$ because the material cleaves well in this direction,
other surfaces are also realized, for example in TI nanoribbons in
the experiment of Ref.~\onlinecite{AharonovBohm}. Our findings are
relevant for those nonstandard surfaces.

Experimentally, the dispersion of surface states is found through
angular resolved photoemission spectroscopy (ARPES)
\cite{Hasan08}. Spin-resolved ARPES provides momentum-resolved
information about the spin polarization of the surface states in
topological insulators
\cite{Hasan09SpinResolved,Roushan09,Rader11}. In
section~\ref{sec:3} we consider the electron's out-of-plane
polarization measured via ARPES, which may be caused by a spin
structure of the surface states that is more complicated than the
spin structure suggested by Eq.~(\ref{surfaceHamiltonian}). The
out-of-plane polarization may arise from the interference effects
of photoelectrons emitted from the two sublattices for surfaces
other than the standard $(111)$ surface, or from photoelectrons
emitted from Se atoms for surfaces that are neither parallel nor
perpendicular to the plane of the quintuple layers."

Another feature of realistic three-dimensional topological
insulators, that is easily overlooked in the effective surface
Hamiltonian (\ref{surfaceHamiltonian}) (but present in the
effective low-energy Hamiltonian of Ref.\ \onlinecite{ZhangBiSe}),
is that the combination of broken particle-hole symmetry and
angular anisotropy of the bulk Hamiltonian leads to different
energies of the Dirac points at different facets of the
topological insulator crystal. Different positions of the Dirac
crossing (or the neutrality point) with respect to Fermi energy
would result in different surface electrons densities at different
surfaces of the crystal. However, an uncompensated charge density
at one or more crystal surfaces comes at a large electrostatic
energetic cost, and one expects a charge redistribution in order
to reduce the Coulomb interaction energy. Below we describe the
solution of the corresponding electrostatic problem at an edge of
a crystal, at the intersection of two differently oriented
surfaces. For short facets, where the Coulomb interaction energies
are not large enough to induce full screening, a finite carrier
density may remain, illustrating the impossibility to tune the
full topological insulator surface to the Dirac point by means of
doping or gating.

\section{Surface states} \label{sec:2}

We first discuss a simplified version of the effective low-energy
continuum Hamiltonian of Ref.\ \onlinecite{ZhangBiSe},
%, so called $k\cdot p$ Hamiltonian,
which consists of a three-dimensional Dirac-like Hamiltonian with
momentum dependent mass,
 \bq\label{HamSymmetric}
  H = \tau_z{\cal M}(\vK) +\tau_x (\vsigma \cdot \vec{\bf K})A, \ \
  {\cal M}(\vK)=M-B K^2,
 \ee
where we use the capital $\vec{\bf K}=({ K}_x,{ K}_y,{ K}_z)$ for
the three-dimensional momentum. The $z$ axis is chosen in the
$(111)$ direction, {\em i.e.}, perpendicular to the plane of the
quintuple layers. The vectors $\vsigma =
(\sigma_x,\sigma_y,\sigma_z)$ and $\vtau = (\tau_x,\tau_y,\tau_z)$
contain two sets of Pauli matrices operating in different
pseudospin spaces. The matrices $\tau_x$, $\tau_y$, and $\tau_z$
refer to the $P1_z^+$ and $P2_z^-$ orbitals of
Refs.~\onlinecite{ZhangBiSe,ModelHam10}, which are such that
%\mbox{Bi} and \mbox{Se} sublattices, such that
states with $\tau_z = 1$ have support mostly on the \mbox{Bi}
sublattice and states with $\tau_z = -1$ have support mostly on
the \mbox{Se} sublattice.
%\cite{ZhangBiSe}.
Below we will refer to these $\tau_z=1$ and $\tau_z=-1$ states
simply as states on the \mbox{Bi} and \mbox{Se} sublattices,
respectively. The electron's spin is expressed in terms of the
Pauli matrices as \cite{FuBerg,ModelHam10}
 \bq\label{spinoperator}
  s_z=\fr{\sigma_z}{2}, \ s_x=\fr{\sigma_x\tau_z}{2}, \
  s_y=\fr{\sigma_y\tau_z}{2}.
 \ee
(The correct form of the spin operator for Bi$_2$Se$_3$ is missing
in Ref.~\onlinecite{ZhangBiSe}.) Equation (\ref{spinoperator})
implies that the operators $\vsigma/2$ and $\vs$ coincide for the
electron residing on the \mbox{Bi} sublattice, whereas the
relation between $\vs$ and $\vsigma/2$ involves a rotation by a
$\pi$ angle around the $z$ axis on the \mbox{Se} sublattice.

The Hamiltonian of Eq.~(\ref{HamSymmetric}) has a higher symmetry
than the true low-energy Hamiltonian of Bi$_2$Se$_3$: It is not
only valence-band--conduction-band (particle--hole) symmetric, but
also symmetric under spatial rotations, provided the Pauli
matrices $\vsigma$ are rotated together with the coordinates,
whereas the Pauli matrices $\vtau$ are kept fixed. The rotation
symmetry is a mathematical artifact of the pseudospin basis used
in Eq.\ (\ref{HamSymmetric}). The strong asymmetry of the real
crystal of Bi$_2$Se$_3$, in which the $z$ axis plays a special
role, is reflected in the anisotropic assignment of the spin
operators $s_x$, $s_y$, and $s_z$ in Eq.\ (\ref{spinoperator}).
Corrections to this simplified model Hamiltonian will be discussed
in Sec.\ \ref{sec:4}. Note that, a direct association of the spin
with $\vsigma$ in Eq.~(\ref{HamSymmetric}) would violate the
parity of the true system. That is why one needs to introduce the
asymmetric spin-operator with the special axis $z$,
Eq.~(\ref{spinoperator}), even in case of the most symmetric model
Hamiltonian~\cite{FuBerg,ModelHam10}.

The Hamiltonian Eq.~(\ref{HamSymmetric}) is of the second order in
spatial derivatives (momentum). Following Ref.\
\onlinecite{ZhangBiSe}, as well as a number of subsequent articles
\cite{FW1,FW2,Lasia2012,NewJPhys10} we require that all four
components of the wavefunction vanish at the crystal surface. This
choice of the boundary conditions guarantees the existence of a
branch of Dirac-like surface states with Dirac crossing at the
$\Gamma$-point (the point of vanishing in-plane momentum), even in
the strongly asymmetric case of Sec.~\ref{sec:4}. We should point
out, however, that this boundary condition is not unique, and that
other choices of boundary conditions have been advocated in the
literature. References~\onlinecite{FuBerg,FuHsieh,Pershoguba12}
suggest to use an effective Hamiltonian that is linear in
momentum, with boundary conditions for which only one pseudospin
component of the wave-function vanishes at the surface. The two
choices of boundary conditions agree qualitatively for the
standard termination at the $(111)$ surface (both give a
Dirac-like branch of surface states near the $\Gamma$ point), but
it is not obvious how to extend the boundary condition of Refs.\
\onlinecite{FuBerg,FuHsieh,Pershoguba12} to surfaces of arbitrary
orientation, and the two boundary conditions may give different
predictions for other surfaces or non-standard surface
terminations \cite{Pershoguba12}.

For the explicit calculation of the surface states from the model
Hamiltonian (\ref{HamSymmetric}) it is convenient to rotate the
coordinate system, such that the TI fills the half-space $z<0$.
This rotation does not change Eq.\ (\ref{HamSymmetric}), but it
changes the relation between the spin operators and the $\vsigma$
matrices: In the rotated coordinate system,
Eq.~(\ref{spinoperator}) takes the form
 \bq\label{spinoperatorN}
  \vs = \fr{(\vsigma\cdot\vn)\vn}{2} + \fr{\vsigma -
(\vsigma\cdot\vn)\vn}{2}\tau_z,
 \ee
where $\vn$ is the unit vector pointing in the direction perpendicular
to the quintuple layer plane in the rotated coordinate frame.

Let us introduce two two-component spinor functions $\psi_B$ and
$\psi_S$ that describe the pseudospin $\tau_z=1$ and $\tau_z=-1$
components of the 4-component wave function. Eigenfunctions of the
Hamiltonian Eq.~(\ref{HamSymmetric}) at energy $\varepsilon$ that
correspond to surface states with momentum $\vec{\bf p}$
%=({ p}_x,{ p}_y)$
parallel to the surface should be found as a linear combinations
of the functions
 \begin{eqnarray}
  \label{surfacewaves}
  \psi_{B} = u_{B} e^{ i(p_x x+p_y y)+\lambda z}, \
  \psi_{S} = u_{S} e^{i(p_x x+p_y y)+\lambda z},
 \end{eqnarray}
with different values of $\lambda$ ($\mbox{Re}\, \lambda > 0$).
Here and below, we use the lower-case symbol $\vec{\bf p}=({
p}_x,{ p}_y,0)$ to denote the in-plane momentum of the surface
states. The spinor amplitudes, $u_B$ and $u_S$, satisfy the
coupled system of equations
 \begin{eqnarray}\label{twoeqsS}
(\eps -{\cal M}(p,\lambda))u_B &=&A(\sigma_x p_x + \sigma_y p_y
-i\lambda\sigma_z)u_S, \nonumber\\
(\eps +{\cal M}(p,\lambda))u_S &=&A(\sigma_x p_x + \sigma_y p_y
-i\lambda\sigma_z)u_B,
 \end{eqnarray}
where ${\cal M}(p,\lambda)=M-Bp^2+B\lambda^2$ and $p^2 = p_x^2 +
p_y^2$. The inverse decay length can take two values,
$\lambda_{1,2}$, which are the solutions of the equation
 \bq\label{energyGeneral}
  \eps^2=(M-Bp^2+B\lambda^2)^2+A^2p^2-A^2\lambda^2.
 \ee
with $\mbox{Re}\, \lambda > 0$. In addition, the surface state
satisfies the boundary condition $\psi_B(z=0)=\psi_S(z=0)=0$, which
provides one additional constraint from which the dispersion
$\varepsilon(p)$ can be calculated.

To find the solution of Eqs.~(\ref{twoeqsS}) and
(\ref{energyGeneral}) one may first guess the (correct) result for
the spectrum,
 \bq\label{cone}
\eps(p)=\pm A p.
 \ee
Then Eq.~(\ref{energyGeneral}) yields a quadratic equation for
$\lambda$,
 \begin{eqnarray}
A \lambda =%\fr{1}{A}
{\cal M}(p,\lambda)=
%\fr{1}{A}
(M-Bp^2+B\lambda^2),
 \end{eqnarray}
where the parameters $A$ and $B$ are taken from the original
Hamiltonian Eq.~(\ref{HamSymmetric}). The first equality here is
the energy of the surface state consistent with the surface
Hamiltonian Eq.~(\ref{surfaceHamiltonian}) with $A = v_{\rm F}$.
The second equality gives the values of
 \bq
%\lambda_{1,2} =\fr{A}{2B} \pm\sqrt{\fr{A^2}{4B^2} +p^2-\fr{M}{B}},
\lambda_{1,2} =({A} \pm\sqrt{{A^2}+{4B^2}p^2-4{M}{B}})/2B,
 \ee
consistent with this value of energy. Substitution of the ansatz
(\ref{cone}) into Eq.~(\ref{twoeqsS}) then gives identical spinors
$u_B$ and $u_S$ for the two roots $\lambda_{1,2}$, which
guarantees that an appropriate linear combination can be found
that satisfies the boundary condition $\psi_B(z=0)=\psi_S(z=0)=0$
for all four components of the spinor wavefunctions $\psi_B$ and
$\psi_S$ simultaneously. The spinor structure corresponding to the
cases $\varepsilon = \pm A p$ is
 \bq\label{uAmplitudes}
  u_B= \frac{1}{2} \left(\begin{array}{cc}
\pm i \\ \gamma
\end{array}\right) \ , \
u_S=\frac{i}{2} \left(\begin{array}{cc} \mp i \\ \gamma
\end{array}\right),
 \ee
with $\gamma = (p_x + i p_y)/p$. The full surface state then has
the form
 \bq
\psi_{B,S}(\vr) = u_{B,S} e^{i(p_x x+p_y y)} (e^{\lambda_1
z}-e^{\lambda_2 z}).
 \ee

For a particle-hole symmetric Hamiltonian Eq.~(\ref{HamSymmetric})
an electron in a surface state is found with equal probability on
either \mbox{Bi} or \mbox{Se} sublattice. In order to clarify the
spin-content of the surface states, we first calculate the
expectation value of the in-plane components of $\vsigma$-operator
on each sublattice. For the in-plane components one finds
\begin{eqnarray}
&& \langle \sigma_i \rangle_B =u_B^\dagger \sigma_i u_B =
  \pm \frac{\eps_{ijz} {p_j}}{{2p}},\nonumber \\
&& \langle \sigma_i \rangle_S = u_S^\dagger \sigma_i u_S = -\langle
\sigma_i \rangle_B ,
\end{eqnarray}
where $i,j=x,y$, whereas
\bq
  \langle \sigma_z \rangle_B = \langle \sigma_z \rangle_S = 0.
\ee
As in Eq.\ (\ref{uAmplitudes}), the upper/lower sign corresponds to
the energies above/below the Dirac crossing. We see that for both
sublattices the expectation value of $\vsigma$ is an
in-surface-plane vector perpendicular to the momentum $p$. However
the direction of $\langle\vsigma\rangle$ is opposite for two
sublattices. As we see from Eqs.~(\ref{spinoperator})
and~(\ref{spinoperatorN}), the direction of $\langle\vsigma\rangle_B$
for the \mbox{Bi} sublattice always coincides with the direction of
the true spin. Thus the spin of this component is always described
by the surface Hamiltonian Eq.~(\ref{surfaceHamiltonian}). On the
contrary, the spin of \mbox{Se} sublattice electron component of the
wave function differs from $\langle \sigma_i \rangle_S$ by a
180-degrees rotation around $\vn$, the axis normal to the layer
plane.

We conclude that the spin orientation of two sublattices coincide
with each other and with the prediction of
Eq.~(\ref{surfaceHamiltonian}) if and only if the surface of the
crystal coincides with the quintuple layer plane. For any other
crystal surface the two sublattices spins differ, and the difference
is typically large. In Fig.~1 we show the spin-directions for two
sublattices for a surface normal to the layer plane.

For certain applications, such as the description of spin and
angle-resolved photoemission (see next Section), it is preferable
to write the Hamiltonian (\ref{HamSymmetric}) in a form in which
the electron spin operator $\vs$ is directly proportional to
$\vsigma$. (Photo-electrons do not carry a sublattice index, so
that a spin-operator that contains $\vtau$ is problematic in that
context.) Hereto, following Ref.~\cite{ModelHam10}, we perform the
unitary transformation
 \bq
  \tilde{H} =  UHU^*, \ U=\fr{1+\tau_z}{2}
  + i(\vsigma\cdot\vn)\fr{1-\tau_z}{2},
 \ee
where $\vn$ is the vector normal to the quintuple layer. The
Hamiltonian $\tilde{H}$ no longer has the manifest rotational symmetry
of Eq.\ (\ref{HamSymmetric}), but the relation between spin and Pauli
matrices now takes the standard form
 \bq\label{spinsimple}
\vs = \vsigma/2.
 \ee
After the unitary transformation the two spinor amplitydes take
the form \bq\label{uTildeAmplitudes}
  \tilde{u}_B =u_B, \ \tilde{u}_S=i(\vsigma\cdot\vn)u_S,
\ee
where $u_B$ and $u_S$ are given in Eq.\ (\ref{uAmplitudes}).

\section{Photo-emission} \label{sec:3}

Spin and angle resolved photoemission spectroscopy (spin-resolved
ARPES) \cite{Dil09} has been the tool of choice for the
experimental investigation of the helical nature of the surface
states in three-dimensional topological insulators
\cite{Hasan09SpinResolved,Roushan09}. What does the coexistence of
two different spin polarizations of the surface electron imply for
the polarization of photoelectrons?

In general, the probability $P$ of electron photoemission is
proportional to the squared matrix element of the interaction
$H_{\rm int}$ with the photon field,
 \bq \label{photoemission}
  P \propto |\langle {\rm f} | H_{\rm int} | {\rm i}\rangle|^2,
 \ee
where $|{\rm f}\rangle$ and $|{\rm i}\rangle$ are the spinor
states corresponding to the free final and the initial surface
electron states, respectively. (One has $H_{\rm int} = (e/c) \vA
\cdot \hat\vj$, where $\vA$ is the vector potential of the photon
field and $\hat\vj$ is the nonrelativistic current density
operator.) In our case, the initial state is described by separate
amplitudes $\tilde u_B$ and $\tilde u_S$ for electrons from the
\mbox{Bi} and \mbox{Se} sublattices, which are given in
Eqs.~(\ref{uAmplitudes},\ref{uTildeAmplitudes}) above. Since the
interaction $H_{\rm int}$ is spin-conserving and local, the spinor
structure $u_{\rm f}$ of the photoelectron becomes a linear
combination
 \bq\label{uf}
u_{\rm f} = \alpha \tilde{u}_B + \beta \tilde{u}_S
 \ee
where the coefficients $\alpha$ and $\beta$ contain
contributions from matrix elements of $H_{\rm int}$ on the \mbox{Bi}
and \mbox{Se} atoms in the crystal, respectively. Their precise value
depends on the details of the atomic structure and the frequency
of the incident light.

Since the surface Hamiltonian
Eq.~(\ref{surfaceHamiltonian}) predicts a spin polarization
parallel to the surface, direct observation of an out-of-plane
spin would be of most interest. A simple calculation gives for the
$z$-component
 \bq\label{interference}
u_{\rm f}^+ \sigma_z u_{\rm f}=
  \frac{(n_z |\beta|^2 + \mbox{Re}\, \alpha\beta^*)(n_x p_y -n_y p_x)}{p}.
 \ee
We see that in case of arbitrary surface orientation the electron
emitted from the \mbox{Se} sublattice acquires an out-of-plane
polarization even without interference. The nature of this
out-of-plane polarization is a direct consequence of the special
form of the spin operators in Eq.~(\ref{spinoperatorN}): The
surface electrons described by the symmetric Hamiltonian
Eq.~(\ref{HamSymmetric}) for each sublattice have the expectation
value of the $\vsigma$-operator parallel to the surface. However,
since the spin of one sublattice is obtained from $\vsigma/2$ by a
$\pi$ rotation around the axis normal to the quintuple layer,
$\vn$, the spins of this sublattice all lie in a plane different
from the crystal surface plane. (Similarly for the asymmetric
Hamiltonian of the following section, for the general surface
orientation, spins of both sublattices at different values of
momentum lie in two planes, different from each other and from the
surface plane.)

As we saw in Sec.~\ref{sec:2}, only for the most studied $(111)$
surface (the surface parallel to the quintuple layer),
polarizations of the electron on both sublattices coincide.
Consequently Eq.~(\ref{interference}) predicts no out-of-plane
polarization here.

Most interesting is the photo-emission from the surface normal to
the quintuple layers ($n_z=0$), where the out-of-plane
polarization appears due to the interference of contributions from
the two sublattices only, $u_{\rm f}^+ \sigma_z u_{\rm f}\sim
\mbox{Re}\, \alpha\beta^*$~(\ref{interference}). The unit cell of
Bi$_2$Se$_3$ consists of 5 atoms, each hosting a $p_z$ electron.
The Hamiltonian Eq.~(\ref{HamSymmetric}) operates in a reduced
$2\times 2$ pseudospin subspace built from an even (with respect
to reflection through the middle plane of the quintuple layer)
state from the conduction band
%(Bi)
and an odd state from the valence band
%(Se)
\cite{ZhangBiSe}. In our case of a surface normal to the quintuple
layer, in the odd state electron waves emitted from the different
atoms would cancel each other, unless there is a finite phase
difference due to the momentum of outgoing electron. Thus we
expect the photoionization amplitude from the odd state to be
$\beta \sim (\vp \cdot \vn)$. Consequently the out-of-plane spin
component Eq.~(\ref{interference}) should have a node and change
sign at $(\vp \cdot \vn) =0$.

\section{Anisotropic topological insulators} \label{sec:4}

In the simplified model of Sec.\ \ref{sec:2}, the origin of the
different spin content of the surface states at different facets
was the special form of the spin operator
Eq.~(\ref{spinoperatorN}). Whereas the special symmetries of the
Hamiltonian Eq.~(\ref{HamSymmetric}), rotational and particle-hole
symmetry, are broken in real crystal, the mechanism leading to the
different spin content of surface states at different facets
continues to operate, as we now show.

The general anisotropic and particle-hole asymmetric
generalization of the Hamiltonian (\ref{HamSymmetric})
is~\cite{Fig2C}
%~\cite{ZhangBiSe}
\bq\label{HamAnisotropic}
  H=\epsilon({\bf K}) +\tau_z{\cal M} +\tau_x\sum_{x,y,z}
A_i\sigma_i {\bf K}_i,
\ee
where now
\bq
  {\cal M}({\bf K})=M-\sum {\bf K}_i B_{ij} {\bf K}_j
\ee
and
\bq
  \epsilon({\bf K})=\sum {\bf K}_i D_{ij} {\bf K}_j,
\ee
with a
positive-definite symmetric matrix $B_{ij}$ and a symmetric matrix
$D_{ij}$. The first term $\epsilon({\bf K})$ in
Eq.~(\ref{HamAnisotropic}) is explicitly particle-hole (valence
band-conduction band) asymmetric.
Choosing the $z$ axis to be orthogonal to the plane of the quintuple
layer of the Bi$_2$Se$_3$ crystal, the matrices $B_{ij}$ and $D_{ij}$
are found to be diagonal \cite{ZhangBiSe}
\begin{eqnarray}
  \label{BDdiagonal}
   B_{ij}=B_i\delta_{ij}, \ D_{ij}=D_i\delta_{ij},
\end{eqnarray}
The microscopic calculations of Ref.~\cite{ZhangBiSe} give the
following values for their elements:
\begin{eqnarray}
  \label{numerical1}
  && A_z=A_1=2.2{\rm eV \AA}, \nonumber \\
  && B_z=B_1=10.{\rm eV \AA}^2, \\
  && D_z=D_1=1.3{\rm eV \AA}^2, \nonumber
\end{eqnarray}
%$  A_z=A_1=2.2{\rm eV \AA},
%A_{x(y)}=A_2=4.1{\rm eV \AA}, B_{zz}=B_1=10.{\rm eV \AA}^2,
%B_{xx(yy)}=B_2=56.6{\rm eV \AA}^2, D_{zz}=D_1=1.3{\rm eV  \AA}^2,
%D_{xx(yy)}=D_2=19.6{\rm eV \AA}^2,  M=0.28{\rm eV} $.
for the direction normal to the layer,
\begin{eqnarray}
  \label{numerical2}
  && A_x=A_y=A_2=4.1{\rm eV \AA}, \nonumber \\
  && B_x=B_y=B_2=56.6{\rm eV \AA}^2,\\
  && D_x=D_y=D_2=19.6{\rm eV \AA}^2, \nonumber
\end{eqnarray}
for the two axes in the layer plane, and
 \bq\label{Mass}
M=0.28{\rm eV}.
 \ee

Finding the surface states for the general Hamiltonian
Eq.~(\ref{HamAnisotropic}) become a complicated algebraic problem.
Here we discuss the solutions only in case that $B_{ij}$ and $D_{ij}$
remain diagonal after a rotation that brings the surface normal to
the $z$ axis. This includes the case of a surface parallel to the
quintuple layer and a surface perpendicular to the layers in
Bi$_2$Se$_3$, as in the nanoribbons of Ref.~\cite{AharonovBohm}.

As in Sec.~\ref{sec:2}, we search for eigenfunctions of the
Hamiltonian (\ref{HamAnisotropic}) of the form (we use again $\vK$
for the three-dimensional momentum and $\vp$ for the
two-dimensional surface states momentum)
 \bq  \psi_{B} = u_{B} e^{ i(p_x x+p_y
y)+\lambda z}, \
  \psi_{S} = u_{S} e^{i(p_x x+p_y y)+\lambda z},
 \ee
where $\mbox{Re}\, \lambda > 0$ and we rotated the coordinate
system, such that the topological insulator occupies the half
space $z < 0$. With the particle-hole asymmetric term
$\epsilon(K)$, simple guessing of the appropriate solution, like
we did in Sec.~\ref{sec:2}, does not work and one has to pursue an
explicit derivation of the result. To this end it is convenient to
replace the sublattice spinors $u_S
=(u_{B_\uparrow},u_{B_\downarrow})$ and $u_S
=(u_{S_\uparrow},u_{S_\downarrow})$ by spin-up and spin-down
pseudospinors
 \bq
  u=(u_{B_\uparrow}, u_{S_\uparrow}), \ \ v=(u_{B_\downarrow},
  u_{S_\downarrow}).
 \ee
These pseudospinor amplitudes now satisfy the system of equations
 \begin{eqnarray}\label{twoeqsIso1}
&(\eps-\epsilon -{\cal M}\tau_z +i\lambda A_z\tau_x)u =(A_xp_x -
iA_y p_y)\tau_x v, ~~~ \\
&(\eps-\epsilon -{\cal M}\tau_z -i\lambda A_z\tau_x)v =(A_xp_x +
iA_y p_y)\tau_x u, ~~~ \label{twoeqsIso2}
 \end{eqnarray}
where
%we abbreviated
 \begin{eqnarray}
  {\cal M}&=&{\cal M}(p,\lambda)=M-B_x p_x^2 -B_y p_y^2
+B_z\lambda^2,\\
  \epsilon&=&\epsilon(p,\lambda)=+D_x p_x^2 +D_y p_y^2
  -D_z\lambda^2.\nonumber
 \end{eqnarray}
{}From the first equation we find
\bq
  \label{uvOmega}
  v=\Omega u,
\ee
with
\bq
  \label{uvOmega2}
  \Omega = \fr{1}{A_xp_x - iA_y p_y}[(\eps-\epsilon)\tau_x
+i{\cal M}\tau_y -i\lambda A_z].
\ee
Substituting this expression for $v$ into Eq.\ (\ref{twoeqsIso2})
one finds
\bq
  \label{AenergyGeneral}
[\eps - \epsilon(p,\lambda)]^2={\cal M}(p,\lambda)^2+A_x^2p_x^2
+A_y^2p_y^2 -A_z^2\lambda^2.
\ee
This equation is biquadratic in $\lambda$ and allows us to find
two values of the squared inverse decay length
$\lambda^2_{1,2}$ for each energy $\eps$, and consequently two
decaying solutions with $\mbox{Re}\, \lambda_{1,2} > 0$.

The open boundary conditions at the surface can be satisfied if
there exists a choice of pseudospinors corresponding to two
solutions $\lambda_1$ and $\lambda_2$ such that
 \bq\label{choice}
  u_1=u_2, \ \ v_1=v_2.
 \ee
The first equality here is easily satisfied, because Eq.\
(\ref{AenergyGeneral}) imposes no restrictions on the pseudospinor
$u$. The second equality then involves the
$\Omega=\Omega(\eps,\lambda)$ of Eq.\ (\ref{uvOmega2}), which
depends on energy and on the specific root of the biquadratic
equation (\ref{AenergyGeneral}),
 \bq
  0=v_1-v_2=[\Omega(\eps,\lambda_1) -\Omega(\eps,\lambda_2)]u.
 \ee
This equation has a solution only if the difference matrix
$\Omega(\eps,\lambda_1) -\Omega(\eps,\lambda_2)$ has a zero eigenvalue,
and consequently $\det [\Omega(\eps,\lambda_1) -\Omega(\eps,\lambda_2)]
=0$. Thus we find the condition
 \bq
\det [(\epsilon_1-\epsilon_2)\tau_x +i({\cal M}_1 -{\cal
M}_2)\tau_2 -iA_z(\lambda_1 -\lambda_2)]=0,
 \ee
where $\lambda_{1}$ and $\lambda_{2}$ are the two roots of
Eq.~(\ref{AenergyGeneral}). This immediately gives a simple
formula
 \bq
A_z^2=(B_z^2-D_z^2)(\lambda_1 +\lambda_2)^2.
 \ee
After straightforward calculation we can now find the energy of
the surface states
 \begin{eqnarray}\label{energyXYZ}
\eps(p) =\fr{D_z}{B_z}M
\pm \sqrt{A_x^2 p_x^2 +A_y^2 p_y^2}\sqrt{1 -\fr{D_z^2}{B_z^2}}\nonumber\\
+\fr{D_xB_z-D_zB_x}{B_z}p_x^2 +\fr{D_yB_z-D_zB_y}{B_z}p_y^2,
 \end{eqnarray}
and the two spinor amplitudes (here we use again the more
meaningful spinor functions for two sublattices instead of the
pseudospinors $u$ and $v$)
%(with $\gamma=(p_x+ip_y)/p$)
\begin{eqnarray}\label{uAmplitudesAnisotropic}
  u_B &=& \fr{1}{2}
  \sqrt{1+\fr{D_z}{B_z}}\left(\begin{array}{cc}
  \pm i \\ \gamma_A \end{array}\right), \nonumber \\
  u_S &=& \fr{i}{2}
  \sqrt{1-\fr{D_z}{B_z}}\left(\begin{array}{cc}
  \mp i \\
  \gamma_A \end{array}\right),
\end{eqnarray}
where now
\bq
  \gamma_A =\sqrt{(A_x p_x+iA_y p_y)/(A_x p_x-iA_y p_y)}.
\ee

The main differences between the solution of the simplified
Hamiltonian of Sec.~\ref{sec:2} and the solution given in
Eqs.~(\ref{energyXYZ}) and (\ref{uAmplitudesAnisotropic}) are:

{\em (i)} The electron no longer spends equal time on the two
sublattices. The difference in probabilities is governed by the
ratio $D_z/B_z$, which can actually be small. (Reference
\onlinecite{ZhangBiSe} finds $D_1/B_1 \approx 0.13$.)

{\em (ii)} For the anisotropic Hamiltonian the expectation value
of the $\vsigma$-operator, which we discussed in
section~\ref{sec:2}, is no longer perpendicular to the momentum.
However, the expectation values $\langle \vsigma\rangle_{B,S}$ on
the Bi and Se sublattices are still perpendicular to the rescaled
momentum $p_x\rightarrow A_x p_x$, $p_y\rightarrow A_y p_y$, and
Fig.~\ref{fig:1} remains valid if the rescaled momentum is used.
We observe that such a rescaling can also be used to account for
the difference between the surface-state spin on the \mbox{Bi} and
\mbox{Se} sublattices that is found here and in Sec.\ \ref{sec:2}:
Rescaling with one inverted sign, $p_x\rightarrow -A_x p_x$,
$p_y\rightarrow A_y p_y$, is sufficient to permute the two panels
in Fig.~\ref{fig:1}.

{\em (iii)} The third important feature of the solution
Eq.~(\ref{energyXYZ}) is also caused by the particle-hole
asymmetric term $\epsilon_0({\vK})$ in Eq.~(\ref{HamAnisotropic}).
As long as one considers the bulk states, this term leads to a
trivial bending of both conduction and valence bands. However, the
inclusion of the energy $\epsilon_0({\vK})$ into the Hamiltonian
(\ref{HamAnisotropic}) leads to a nonzero value of the Dirac
crossing energy, as can be seen from the explicit solution
(\ref{energyXYZ}). In the most general case of matrices $D_{ij}$
and $B_{ij}$ having no vanishing elements we found the position of
the Dirac crossing, which is the (unique) energy eigenvalue of a
surface state with vanishing momentum parallel to the surface, to
be simply $\varepsilon =\eps_{\rm Dirac}=M D_{zz}/B_{zz}$. Thus,
returning to the original coordinate system of
Eqs.~(\ref{HamAnisotropic})-(\ref{Mass}), we may introduce a
vector $\ve$ normal to the surface at a particular point and write
a formula for the energy of the Dirac crossing valid for the
entire crystal
 \bq\label{DiracGen}
\eps_{\rm Dirac}= M({\sum {\bf e}_i D_{ij} {\bf e}_j})/({\sum {\bf
e}_i B_{ij} {\bf e}_j}).
 \ee
Substituting the actual values of the matrices $D_{ij}$ and $B_{ij}$,
we find that for Bi$_2$Se$_3$ the position of the Dirac crossing may
vary by as much as
\bq\label{DiracRange}
  \Delta\eps_{\rm Dirac}= \left( \fr{D_2}{B_2}
  -\fr{D_1}{B_1}\right)M\approx .06{\rm eV} \approx \fr{A_2}{6.8
  {\rm nm}},
\ee
upon varying the orientation of the surface.

\begin{figure}
\includegraphics[width=4.5cm]{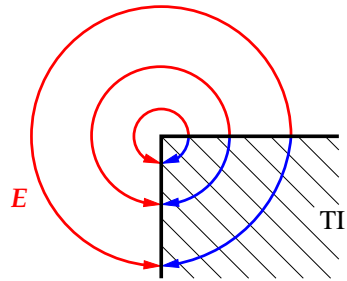}
\caption{\label{fig:2} Electric field near the edge of an
anisotropic TI.}
%\vspace{-.3cm}
\end{figure}

\section{Electrostatics of a topological insulator edge} \label{sec:5}

Being built of neutral atoms, the topological insulator crystal is
obviously a charge-neutral system. Because of the bulk band gap,
the bulk charge density is insensitive to the precise choice of
the Fermi energy. However, the requirement of neutrality should
also include the surface charges. Even in the absence of
particle-hole symmetry, the structure of the surface-state
spectrum is such, that generically charge neutrality is achieved
if the Fermi energy equals the energy of the Dirac crossing,
$\varepsilon_{\rm Dirac}$. Since the latter energy depends on the
orientation of the surface, see Eqs.~(\ref{DiracGen}) and
(\ref{DiracRange}), surface-charge neutrality can not be satisfied
by a uniform choice of the Fermi energy for the entire crystal.
(The variation of the surface charge density for a hypothetical
uniform choice of the Fermi energy is estimated as $\Delta n\sim
(\Delta\eps_{\rm Dirac}/A_2)^2/4\pi \approx 1.7\cdot 10^{11}{\rm
cm}^{-2}$.) Such a choice of the Fermi energy would lead to a
strong electric field along the surface, which should not happen
for a metallic surface. Instead, the surface charge is
redistributed near the edges between different facets, so that the
electric field parallel to the metallic surface vanishes. Far away
from the edge, surface neutrality will then be achieved
simultaneously for both surfaces, while the electrostatic
potential difference between two facets will compensate the
difference of the Dirac crossing energies, $\Delta\eps_{\rm
Dirac}= \Delta V$. In the limit that the surfaces can be
approximated as perfect metals, the solution of such electrostatic
problem is straightforward \cite{LandauECM}. For two surfaces at a
90 degree angle, as shown in Fig.~\ref{fig:2}, the potential
$\Phi_{\rm in}$ and $\Phi_{\rm out}$ inside (polar angle
$-\pi/2<\phi<0$) or outside (polar angle $0 < \phi < 3 \pi/2)$ the
topological insulator is
 \bq\label{Phi}
\Phi_{\rm out}(r,\phi)=\fr{2\Delta V}{3\pi e} \phi, \  \ \Phi_{\rm
in}(r,\phi)=-\fr{2\Delta V}{\pi e} \phi,
 \ee
corresponding to a surface charge density
 \bq\label{chargeV}
|\sigma(r)| = (\kappa +\fr{1}{3})\fr{\Delta V}{2\pi^2 e r}
% \approx \kappa \fr{e}{r}\times 3\cdot 10^{4}{\rm cm}^{-1}.
 \ee
at a distance $r$ away from the edge. Substituting the dielectric
constant of Bi$_2$Se$_3$, $\kappa \approx 110$, and taking $\Delta
V =.06{\rm eV}$ from Eq.~(\ref{DiracRange}), we arrive at the
estimate $|\sigma(r)| \sim  (e/r) \times 2.3\cdot 10^{6}{\rm
cm}^{-1}$. Doping of the surface electron gas by electrostatic
external gates would lead to an additional edge/corner charge
accumulation, similar to the edge charge accumulation in graphene
ribbons~\cite{Sil08}.

The electrostatic calculation of Eqs.~(\ref{Phi},\ref{chargeV})
ignores completely the kinetic energy of the surface electrons.
However, due to the large dielectric constant in Bi$_2$Se$_3$
there exists a regime where the semiclassical electron dynamics is
already described by a smooth coordinate dependent Fermi energy
$E_F(r)= A k_F(r)$, but the electrostatic energy is not yet
dominant, $e\Phi\sim E_F$ (Thomas Fermi approximation). In this
case one requires the sum $E_F(r)+e\Phi $ to be a constant over
the metallic surface, while electrostatics is recovered in the
large-sample limit.  The range of validity of Eq.~(\ref{Phi}) is
now found as $\Delta V> A_2 k_F(r)=A_2 2\sqrt{\pi\sigma(r)/ e} $,
leading to $r> 135 {\rm nm}$. For much smaller samples one should
ignore the Coulomb interaction and describe the surface electrons
by the Dirac equation with shifted crossing energy,
Eqs.~(\ref{DiracGen},\ref{DiracRange}).

\section{Discussion} \label{sec:6}

In this article, we have analyzed the spin structure of surface
states in the three-dimensional topological insulator
Bi$_2$Se$_3$, as it follows from the low-energy continuum
description proposed in Ref.\ \onlinecite{ZhangBiSe}. The
inequivalence of the spin variables on the Bi and Se sublattices
in this continuum description leads to a nontrivial spin structure
of the surface states at surfaces other than the $(111)$ surface.
We also found that the energy of the Dirac crossing in the
surface-state dispersion depends on the surface orientation.
Although the precise boundary conditions of the effective
low-energy description --- open boundary conditions, in which all
components of the spinor wavefunction vanish at the surface of the
topological insulator, as in Ref.\ \onlinecite{ZhangBiSe} --- are
key to our quantitative analysis, we expect that the effects we
predict persists if the boundary conditions are changed. Different
boundary conditions have appeared in the literature for the
$(111)$ surface \cite{FuBerg,FuHsieh,Pershoguba12}, but not for
other surfaces of a Bi$_2$Se$_3$ crystal.

The theoretical findings of this paper may be
applied to the interpretation of several experiments involving the
surface states of three-dimensional topological insulators. Here
we mention the STM measurements of the surface electrons charge
near an artificial step on Bi$_2$Te$_3$
surface~\cite{Kapitulnik09,Kapitulnik10} and the measurement of
Aharonov-Bohm oscillations in the surface-state-mediated transport
in Bi$_2$Se$_3$ nanoribbons~\cite{AharonovBohm}. In the latter case
a theoretical analysis predicts that the magnetoconductance should
depend strongly on the position of the Dirac crossing for the
electrons on the surface of the nanoribbon \cite{Bardarson2010}.
Obviously, a non-uniformity of this Dirac crossing of the type
discussed above will affect the minimal conductivity and
interference pattern for the thick ($\sim 100{\rm nm}$)
rectangular shaped nanoribbons of Ref.~\cite{AharonovBohm} and
should be taken into account in a quantitative modeling of the
device.

Our most remarkable result is the prediction of the possibility of
out-of-plane momentum polarizations of photo-emitted electrons,
Eq.~(\ref{interference}), depending on the exact orientation of
the surface. Measuring these spin components would confirm the
validity of the microscopic Hamiltonian of Ref.~\cite{ZhangBiSe}
and its boundary conditions. It will also explicitly demonstrate
the signatures of interference of photoemission contributions from
two sublattices of Bi$_2$Se$_3$.
%, and be a strong signature of interference of photoemission
%contributions from both sublattices of Bi$_2$Se$_3$.
Such experiments require, however, the preparation of Bi$_2$Se$_3$
crystals with sufficient quality surfaces other than in the
$(111)$ direction.
%, something
%that has not yet been achieved with a sufficient quality that
%spin-resolved ARPES measurements are possible.

%\\
% {\bf To change the last sent., especially if we found 110 of
%Hasan.}

Although the low-energy model of Ref.\ \onlinecite{ZhangBiSe},
that we used as the basis for our calculations, predicts a
nontrivial spin structure for surfaces other than the $(111)$
surface, there may be other mechanisms that lead to an effectively
reduced spin that are not included in this model. In this context,
we mention recent first-principles calculations of Yazyev and
coworkers, who find that the net spin polarization of surface
states on the $(111)$ surface is reduced by an amount of order
50\% \cite{Louie2010}, which they
%the authors of Ref.\ \onlinecite{Louie2010}
attribute to the effect of strong spin-orbit interaction on the
heavy Bi atoms. Since the magnitude of the spin of a photo-emitted
electron has to be $s=1/2$ (a free electron always has its spin
pointed in some direction), the reduced net spin found in Ref.\
\onlinecite{Louie2010} implies a likely out-of-plane component of
photo-emitted electrons, even for the $(111)$ surface. This effect
may be suppressed since it involves the photo-ionization from the
inner row of Bi atoms, less accessible for ARPES. On the other
hand, as the spin-orbit interaction admixes $p_x\pm ip_y$ electron
states to $p_z$ ones, one may play with the photon polarization to
selectively enhance the electron's ionization from Bi. Further
investigation in this direction is obviously desired.

%{\bf old:} , with a mechanism similar to that discussed in Sec.\
%\ref{sec:3}.

\acknowledgements

Discussions with Ewelina Hankiewicz, Gene Mele, Felix von Oppen,
Oliver Rader, Andrei Varykhalov, and Gergely Zarand  as well as
correspondence with Liang Fu and Shou-Cheng Zhang are greatly
appreciated. This work is supported by the SFB TR 12,
%of the German Science Foundation
%and
by the Alexander von Humboldt Foundation and  by DOE, Office of
Basic Energy Sciences, Grant No.~DE-FG02-06ER46313~(E.M.).

{\it Note added --} Upon completing this version of the article,
we learned of Ref.~\cite{ZhangKaneMele12}, which also addresses
the spin structure and Dirac crossing of the surface states.

%When we were completing this revised version of the paper, we
%became aware Ref.~\cite{ZhangKaneMele12}, where authors
%%of Ref.~\cite{ZhangKaneMele12}
%consider the face-orientation-dependent energy of the Dirac
%crossing for the surface states, consistent with our result. The
%surface spin textures considered by the authors correspond to the
%over the pseudospin componets spin of the electron, the quantity
%which is not measured in the experiment.

\end{document}